\documentclass[aps,prd,twocolumn,showpacs]{revtex4}

\usepackage{graphicx}
\usepackage{bbm}

\newcommand{\nn}{\nonumber\\}

\begin{document}

\title{A mass-dependent $\beta$-function}

\author{Dennis D.~Dietrich}

\affiliation{CP$^3$-Origins, Centre for Particle Physics Phenomenology, University of Southern Denmark, Odense, Denmark}

\date{September 14, 2009}

\begin{abstract}

Threshold effects related to fermion masses are considered for an all-order $\beta$-function based on a background field momentum subtraction scheme. Far away from all thresholds, the suggested $\beta$-function reduces to the conjectured all-order form inspired by the Novikov--Shifman--Vainshtein--Zakharov $\beta$-function of $\mathcal{N}$=1 supersymmetric gauge theories with a fixed integer number of fermion flavours. At (formally) infinite masses the corresponding pure Yang--Mills $\beta$-function is recovered. We discuss applications to the phase diagram of non-Abelian field theories.

\end{abstract}


\maketitle


\section{Introduction}

The $\beta$-function of non-Abelian gauge field theories is scheme independent (universal) up to two loops. This statement holds if a fixed number of active matter species is considered. Regarding the renormalisation group running of the coupling for a fixed number of flavours is legitimate as long as the energy scales under consideration are either far above or below the fermion masses, that is, the relevant thresholds. From the viewpoint of renormalisation theory, it is legal to use the fixed flavour schemes even close or across the thresholds, although the freezing out (By default, we evolve from higher to lower energy scales.) of heavy flavours is ignored. Sometimes this effect is incorporated by piecing together $\beta$-functions for different integer numbers of flavours exactly at the mass of the flavours which become inactive. While this procedure might appear somewhat ad hoc, it features already a crucial aspect of the threshold effect, the scale dependence of the $\beta$-function coefficients. They are no longer scheme independent in the above sense. The switching also involves a blending between different schemes. It is plausible that the transition between different numbers of active flavours at a threshold should happen gradually. 
Therefore, we base the following investigation on a background field momentum subtraction scheme \cite{Jegerlehner:1998zg}, which features naturally a smooth switching. Momentum subtraction schemes \cite{De Rujula:1976au,Georgi:1976ve,Nachtmann:1978vf,Shirkov:1992pc,Chyla:1995fm} respect the decoupling theorem \cite{Appelquist:1974tg}, but, in general, spoil the Slavnov--Taylor identities. The latter shortcoming is cured by the use of the background field method \cite{DeWitt:1980jv}. Alternatively, we could have used the physical charge approach of Ref.~\cite{Brodsky:1999fr}. The outcome, however, is qualitatively and quantitatively close to that in the background field momentum subtraction scheme (See especially Fig.~2 in \cite{Brodsky:1999fr}.) and there the expressions can be handled analytically.

An all-order $\beta$-function without threshold effects has been conjectured in \cite{Ryttov:2007cx}. It is inspired by the Novikov--Shifman--Vainshtein--Zakharov $\beta$-function of $\mathcal{N}$=1 supersymmetric gauge theories \cite{Novikov:1983uc}. The conjectured $\beta$-function possesses the correct limits for exactly known cases, like super Yang--Mills theory \cite{Novikov:1983uc} or planar equivalence in a large-$N_c$ limit \cite{Armoni:2003gp}. At two-loop order it coincides with the universal $\beta$-function. 

We base our suggestion for a mass dependent all-orders $\beta$-function on the postulate that, far away from all thresholds, it reduces to the just mentioned all-order $\beta$-function for a fixed number of flavours. At two-loop order it is to be identical with the $\beta$-function in background field momentum subtraction scheme \cite{Jegerlehner:1998zg}. From the thus obtained mass-dependent $\beta$-function we find that threshold effects are felt two orders of magnitude away from the mass of the fermion. (See Fig.~\ref{mwt}.)

\begin{figure}[t]
 \resizebox{8.5cm}{!}{\includegraphics{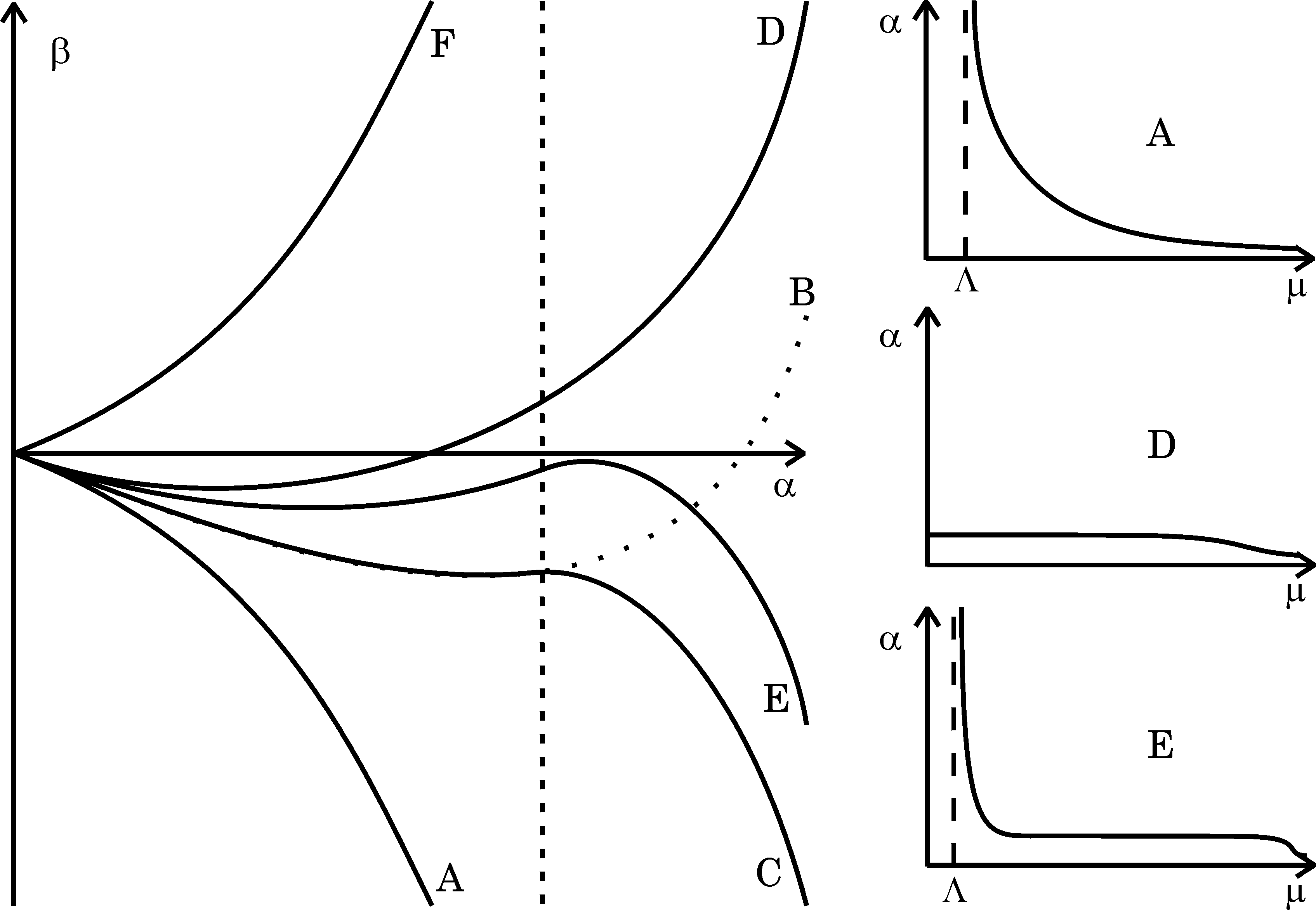}}
 \caption{Behaviour of the $\beta$-function as a function of the coupling $\alpha$ and of the coupling as a function of the energy scale $\mu$, depending on the matter content of the theory.
 A) No or little matter;
 B) existence of a perturbative Caswell--Banks--Zaks fixed point;
 C) actual shape due to chiral symmetry breaking;
 D) realised fixed point;
 E) quasiconformal case;
 F) loss of asymptotic freedom.
 The dashed line in the plot on the left-hand side indicates the critical values of the coupling for chiral symmetry breaking.}
 \label{beta}
 \end{figure}

One field of research where the mass-dependence of the $\beta$-function is of importance is the conformal window of non-Abelian gauge field theories. In the picture laid out in \cite{Dietrich:2006cm}, the interplay between the matter content of a theory and chiral symmetry breaking gives rise to different phases as follows (See Fig.~\ref{beta}.): For no or only little matter the (dominant) antiscreening of the non-Abelian gauge bosons inhibits the occurrence of an infrared fixed point (A). This is the case for quantum chromodynamics. For slightly more matter a perturbative Caswell--Banks--Zaks \cite{Caswell:1974gg} fixed point develops. When arguing based on the two-loop $\beta$-function, this amounts to a change of the sign of the second coefficient (B). This fixed point, however, need not be realised, as, for an insufficient matter content, the value of the coupling may suffice to trigger chiral symmetry breaking. In that case, the fermions acquire a dynamical mass and decouple, at least in parts, from the dynamics. The effective number of flavours is reduced and the antiscreening dominates again (C). Only above a certain amount of matter, the fixed point is reached before chiral symmetry breaking sets in. The coupling constant freezes (D). Just before this, the quasiconformal case is to be found, where the fixed point is almost realised (E). There the $\beta$-function is small for a value of the coupling $\alpha$ close to its critical value for chiral symmetry breaking. Consequently, the coupling stays almost constant (it walks) for a large interval of scales before chiral symmetry breaking is triggered and the coupling constant starts running again. Beyond a given amount of matter the theory loses asymptotic freedom. The first coefficient of the $\beta$-function changes sign. This discussion indicates why the acquisition of (more) mass by the fermions is a crucial building block for the understanding of the dynamics of the theory. 

The motivation for carrying out the investigation in \cite{Dietrich:2006cm} was the determination of walking, that is, quasiconformal \cite{walk} technicolour \cite{TC} models, which are consistent with available electroweak precision data \cite{Dietrich:2005jn,Dietrich:2005wk}. \{Walking technicolour models possess a rich collider phenomenology for the Large Hadron Collider (LHC) \cite{Belyaev:2008yj} as well as dark matter candidates \cite{Gudnason:2006ug} and are also interesting models for studies in the AdS/CFT framework \cite{Hong:2006si}.\} In technicolour models, chiral symmetry breaking among fermions (techniquarks) added to the standard model without Higgs sector simultaneously breaks the electroweak symmetry dynamically. 
Technicolour hence provides the masses for the weak gauge bosons. The most intensely researched way to also render massive the standard model fermions is extended technicolour \cite{ETC}. Apart from this primary purpose, extended technicolour may serve to make extra Nambu--Goldstone-modes (Three of the potentially more modes arising from the chiral symmetry breaking are absorbed as the longitudinal modes of the weak gauge bosons.) sufficiently massive, and or stabilise the vacuum alignment \cite{Peskin:1980gc,Dietrich:2009ix}. The effects from extended technicolour are already in place before the chiral condensate of technicolour forms. Therefore, as shall be explained below, some of the effects connected to the stabilisation of the vacuum alignment and the Nambu--Goldstone masses are similar in nature to an explicit mass term in the sense of the electroweakly induced quark masses relative to the chiral dynamics of quantum chromodynamics.

The just described occurrence of walking is due to inherently non-perturbative effects from chiral symmetry breaking. In the context of the mass-dependent $\beta$-function discussed here, we would like to point out that there is also the possibility to see walking from an interplay of more perturbative effects: Let us think of an asymptotically free gauge theory, which would have an infrared fixed point for massless fermions. ``Hard" fermion masses in the sense of, for example, the electroweakly induced quark masses as seen by quantum chromodynamics lead to a reduction of the screening due to the fermions. If this occurs at the correct energy scale (Here, the occurrence of walking depends on the initial condition for the coupling constant.), the infrared fixed point is never reached, but a situation analogous to the one depicted in Fig.~\ref{beta}E can arise.

The paper is organised as follows. In Sect.~\ref{BETA}, we derive our mass-dependent all-order $\beta$-function. Sect.~\ref{CONF} contains the application to the question of the determination of the conformal window.  
Sect.~\ref{CONF2} studies the quantitative influence of fermion masses on the (quasi)conformal window. Sect.~\ref{SUMM} summarises the results.
The Appendix contains plots of the (quasi)conformal windows for gauge field theories with fermions transforming under various representations of $SU(N_c)$, $Sp(2N_c)$ and $SO(N_c)$ gauge groups.


\section{A mass dependent $\beta$-function\label{BETA}}

The $\beta$-function describes the change of the gauge coupling $g$ of non-Abelian gauge field theories with the energy scale $\mu$. It is subtraction scheme independent up to two loops,
\begin{equation}
\beta (g)=-\frac{\beta_0}{(4\pi)^2}g^3-\frac{\beta_1}{(4\pi)^4}g^5-\dots ,
\end{equation}

\begin{equation}
\beta_0=\frac{11}{3}C_2(G)-\frac{4}{3}T(R)N_f ,
\end{equation}

\begin{equation}
\beta_1=\frac{34}{3}C_2(G)^2-\frac{20}{3}C_2(G)T(R)N_f-4C_2(R)T(R)N_f .
\end{equation}
In the background field momentum subtraction scheme, however, the first two $\beta$-function coefficients are different \cite{Jegerlehner:1998zg},
\begin{eqnarray}
\beta_0\mapsto\bar{\beta}_0&=&\frac{11}{3}C_2(G)-\frac{4}{3}T(R)\sum_{j=1}^{N_f}b_0(x_j) ,\label{barbeta0}\\
\beta_1\mapsto\bar{\beta}_1&=&\frac{34}{3}C_2(G)^2-T(R)\sum_{j=1}^{N_f}b_1(x_j) .
\label{barbeta1}
\end{eqnarray}
Here,
$
x_j=-\mu^2/(4m_j^2),
$
$m_j$ is the mass of the fermion flavour $j$. Further,
\begin{equation}
b_0(x)=1+3[1-G(x)]/(2x) ,
\end{equation}
where
$
G(x)=(2y\ln y)/(y^2-1)
$
as well as
$
y=(\sqrt{1-1/x}-1)/(\sqrt{1-1/x}+1)
$
and
\begin{widetext}
\begin{eqnarray}
b_1(x)
&=&
\frac{16(1-x^2)C_2(R)+(1+8x^2)C_2(G)}{6x^2(1-x)}\sigma(x)
-
\frac{2}{3x^2}(C_2(G)-2C_2(R))I(x)
+
\frac{2}{3x}\tilde I_3^{(4)}(x)C_2(G)
+\nn&&+
[(1+3x-10x^2+12x^3)C_2(G)-3(3-3x-4x^2+8x^3)C_2(R)]\frac{4}{3x}G(x)^2
-\nn&&-
[(147-4x-100x^2+8x^3)C_2(G)+168(1-x)C_2(R)+6(9+4x)\ln(-4x)C_2(G)]\frac{1}{9x}G(x)
+\nn&&+
[(99+62x)C_2(G)+12(11+3x)C_2(R)+2(27+24x-2x^2)\ln(-4x)C_2(G)]\frac{1}{9x} ,
\end{eqnarray}
\begin{equation}
\sigma(x)
=
\left\{2\mathrm{Li}_2(-y)+\mathrm{Li}_2(y)+[\ln(1-y)+2\ln(1+y)-(3/4)\ln y]\ln y\right\}
(1-y^2)/y 
\end{equation}
\begin{equation}
I(x)
=
6[\zeta_3+4\mathrm{Li}_3(-y)+2\mathrm{Li}_3(y)]-8[2\mathrm{Li}_2(-y)+\mathrm{Li}_2(y)]\ln y-2[2\ln(1+y)+\ln(1-y)]\ln^2y ,
\end{equation}
\begin{equation}
\tilde I^{(4)}_3(x)
=
6\zeta_3-6\mathrm{Li}_3(y)+6\mathrm{Li}_2(y)\ln y+2\ln(1-y)\ln^2y .
\end{equation}
\end{widetext}
Here, $\zeta_3=\zeta(3)=1.2020569\dots$ stands for Ap\'ery's constant, that is, Riemann's zeta function evaluated at 3.
$\mathrm{Li}_n(z)=\sum_{j=1}^\infty\frac{z^j}{j^n}$
is the polylogarithm.

In the background field method, the $\beta$-function is obtained from the background field renormalisation constant $Z_A$ according to,
\begin{equation}
\mu^2\frac{d}{d\mu^2}\frac{\alpha}{4\pi}
=
\lim_{\epsilon\rightarrow 0}\alpha\mu\frac{d}{d\mu}\ln Z_A .
\label{bfbeta}
\end{equation}
Here, $\alpha=g/(4\pi)$ and $\epsilon$ emanates from dimensional regularisation, $d=4-2\epsilon$. (Complications arising from the renormalisation of the gauge parameter can be avoided by adopting background field Landau gauge, but the physical result does not depend on this choice.) Consequently, only background field propagator diagrams have to be evaluated, that is, diagrams with two external couplings to the external field and no other external legs. $Z_A$ is obtained from,
\begin{equation}
Z_A=\frac{1+\Pi_0(Q^2,\mu^2,\{m_j^2\})}{1+\Pi(Q^2,\mu^2,\{m_j^2\})},
\end{equation}
where $\Pi_0$ is the bare and $\Pi$ the renormalised self-energy amplitude. The renormalised masses $\{m_j\}$ of the fermions be defined as poles of the corresponding propagators. In the momentum subtraction scheme, the renormalised self-energy amplitude is fixed by the condition,
\begin{equation}
\Pi(Q^2,\mu^2,\{m_j^2\})|_{Q^2=\mu^2}=0 .
\end{equation}
After computing the renormalised self-energy amplitude to two-loop order, the corresponding $\beta$-function can be extracted by differentiation according to Eq.~(\ref{bfbeta}). In this context, $I(x)$ and $\tilde I^{(4)}_3(x)$ are master integrals appearing in the computation of $\Pi$.

Let us consider $N_f$ mass degenerate flavours, all transforming under the same representation of the gauge group. Then the modifications of $\bar{\beta}_0$ and $\bar{\beta}_1$ relative to $\beta_0$ and $\beta_1$ can be collected in ``numbers of active flavours",
\begin{eqnarray}
N_{f,0}/N_f&=&b_0(x) ,\\
N_{f,1}/N_f&=&b_1(x)/[(20/3)C_2(G)+4C_2(R)] .
\end{eqnarray}
Both go to $N_f$ for massless fermions and to zero for very massive fermions, thereby satisfying the decoupling theorem \cite{Appelquist:1974tg}. The concept of number of active flavours can even be given a gauge invariant meaning \cite{Brodsky:1999fr}, which makes such numbers observable, in principle. This, however, raises all the more the question why there should be different numbers of active flavours in each term of the $\beta$-function. Furthermore, $N_{f,0}$ interpolates monotonously between the two limiting cases $N_f$ and zero, which would be expected from a number of active flavours, while $N_{f,1}$ does not. The latter is, in general, not even positive for all values of $x$. This suggests that the interpretation as active number of flavours is more appropriate for $N_{f,0}$ than it is for $N_{f,1}$. We shall revisit this point below.

The apparent contradiction to the statement about scheme independence of the two-loop $\beta$-function arises from the fact that the background field momentum subtraction scheme is not a fixed subtraction scheme in the above sense, but can be seen as a smooth interpolation between subtraction schemes as a function of the scale \cite{Shirkov:1992pc}. The transition takes place when the scale closes in on the mass of a flavour, which then freezes out gradually. 
We can also think about the other way to incorporate the change of the number of active flavours with the energy scale, that is, the matching of running couplings for different integer numbers of flavours. The matching is performed exactly where the scale meets the mass of a given flavour. In this sense, also the two-loop $\beta$-function coefficients are functions of the scale and do not coincide with the subtraction scheme independent expressions for a fixed number of flavours.

Based on the argument of universality in \cite{Ryttov:2007cx}, an all-order $\beta$-function was conjectured,
\begin{equation}
\beta(g)
=
-
\frac{g^3}{(4\pi)^2}
\frac{\beta_0-\frac{2}{3}T(R)N_f\gamma(g^2)}{1-\frac{g^2}{8\pi^2}C_2(G)(1+2\frac{\beta^\prime_0}{\beta_0})} ,
\label{fsaob}
\end{equation}
where
$
\beta^\prime_0=C_2(G)-T(R)N_f .
$
$\gamma=-d\ln m/d\ln\mu$
stands for the anomalous dimension of the fermion mass operator,
$
\gamma(g^2)=(3/2)C_2(R)g^2/(4\pi^2)+O(g^4).
$
An expansion to two-loop order reproduces the universal two-loop coefficients.
Many expression would achieve this, but, as shown in \cite{Ryttov:2007cx}, (\ref{fsaob}) is consistent with limiting cases in which the $\beta$-function is known exactly like super Yang--Mills theory \cite{Novikov:1983uc} or planar equivalence in a large-$N_c$ limit \cite{Armoni:2003gp}.

Our objective is to find a mass dependent all-order $\beta$-function that when expanded to second order reproduced the two-loop $\beta$-function in the background field momentum subtraction scheme. (If we chose a different mass-dependent two-loop $\beta$-function as starting point the resulting all-order $\beta$-function would be accordingly different.) Further, in the massless limit the mass dependent all-order $\beta$-function is to coincide with the mass-independent all-order $\beta$-function. This ascertains that the exactly known supersymmetric results can be reproduced. Likewise, in the ultramassive limit it is to coincide with the pure Yang--Mills version of the mass-independent all-order $\beta$-function. This implies also that all terms involving the Casimir $C_2(R)$ have to be absorbed in the term involving the anomalous dimension.

We propose the following mass dependent $\beta$-function,
\begin{equation}
\bar{\beta}(g)
=
-
\frac{g^3}{(4\pi)^2}
\frac{\bar{\beta}_0-\frac{2}{3}T(R)\sum_{j=1}^{N_f}\bar{\gamma}(x_j)}{1-\frac{g^2}{8\pi^2}C_2(G)(1+2\frac{\bar{\beta}^\prime_0}{\bar{\beta}_0})} ,
\label{maob}
\end{equation}
where
\begin{equation}
\bar{\gamma}(x_j)=
{\frac{3}{2}\frac{g^2}{4\pi^2}}\frac{1}{4}b_1(x_j)|_{C_2(G)\rightarrow 0}
\end{equation}
and
\[
\bar{\beta}^\prime_0
=
C_2(G)
+
T(R)\sum_{j=1}^{N_f}\left[
\frac{2}{3}b_0(x_j)
-
{
\frac{1}{4}\frac{b_1(x_j)|_{C_2(R)\rightarrow 0}}{C_2(G)}
}
\right] .
\]
It fulfils all of the aforementioned requirements:
Taking all masses $m_j$ to zero reproduces Eq.~(\ref{fsaob}). In fact, all barred, that is, mass dependent quantities $\bar{\beta}_0$, $\bar{\beta}_1$, $\bar{\beta}_0^\prime$ and especially, $\bar{\gamma}$ go to their unbarred, that is, massless counterparts separately. $\bar{\gamma}(x_j)/\gamma$ has the same limits, 0 for $x\rightarrow 0$ and 1 for $x\rightarrow-\infty$, as found from the mass dependent anomalous dimension $\gamma_m=\gamma_{m=0}[1+\ln(1-4x)/(4x)]$ given in \cite{Georgi:1976ve}. 
Expanding to two-loop order reproduces Eqs.~(\ref{barbeta0}) and (\ref{barbeta1}).
[This last feature would also be achieved if a factor $1+O(g^2)$ was introduced in $\bar{\gamma}$ and/or $\bar{\beta}_0^\prime$. (The same is already true in the massless case.) Its effect would, however, only make appearance at third order and could, hence, be absorbed in a change of the renormalisation scheme. Likewise, if one wanted to accommodate a particular three-loop term, one could include such a factor and adjust the denominator accordingly, which would amount to a change of scheme.]

Like Eq.~(\ref{fsaob}), Eq.~(\ref{maob}) can be generalised to allow for flavours which transform under different representations of the gauge group. In this case it reads,
\begin{equation}
\bar{\beta}(g)
=
-
\frac{g^3}{(4\pi)^2}
\frac{\bar{\beta}_0-\frac{2}{3}\sum_{j=1}^{N_f}T(R_j)\bar{\gamma}(x_j)}{1-\frac{g^2}{8\pi^2}C_2(G)(1+\frac{\bar{\beta}^\prime_0}{\bar{\beta}_0})} ,
\end{equation}
where
\begin{equation}
\bar{\beta}^\prime_0=C_2(G)+\sum_{j=1}^{N_f}T(R_j)\left[\frac{2}{3}b_0(x_j)-
{
\frac{1}{4}\frac{b_1(x_j)|_{C_2(R)\rightarrow 0}}{C_2(G)}
}\right]
\end{equation}
and $R_j$ is the representation of flavour $j$.

Through Eq.~(\ref{maob}), the coupling $g$ now depends on the anomalous dimension $\gamma$ in a twofold way: As in Eq.~(\ref{fsaob}), there is a direct dependence in the numerator. Here additionally, the $\beta$-function depends on the masses, which via the renormalisation group equations
\begin{eqnarray}
\mu\frac{d}{d\mu}g(\mu)
&=&
+\beta[g(\mu),\{m_k(\mu)/\mu\}],\\
\mu\frac{d}{d\mu}m_j(\mu)
&=&
-\gamma[g(\mu),\{m_k(\mu)/\mu\}] ,
\end{eqnarray}
are functionals of the anomalous dimension.


\section{Impact on the conformal window\label{CONF}}

Here, predominantly for the sake of notational simplicity, we concentrate on walking technicolour models with two techniflavours, the generalisation to more flavours being straightforward. (The most intensely studied technicolour models, minimal and next-to-minimal walking technicolour \cite{Sannino:2004qp,Dietrich:2005jn,Dietrich:2005wk}, feature two techniflavours.)

Models in which the techniquarks do not transform under a (pseudo)real representation of the technicolour gauge group have the essential $SU(2)_L\times SU(2)_R$ flavour symmetry. When it breaks to $SU(2)_V$, the electroweak $SU(2)_L\times U(1)_Y$ breaks to $U(1)_\mathrm{em}$, regardless of the embedding \cite{Peskin:1980gc}. The breaking leads to three pseudo-Nambu--Goldstone-modes, which become the longitudinal degrees of freedom of the weak gauge bosons.

Models with techniquarks in (pseudo)real representations of the technicolour gauge group have an $SU(4)$ unbroken flavour symmetry. For real representations it breaks to $SO(4)$, which leaves behind six Nambu--Goldstone-modes beyond the three degrees of freedom, which are absorbed as longitudinal degrees of freedom of the weak gauge bosons. They have to be sufficiently massive, as these potentially lightest states of a technicolour theory have not been detected to date. It turns out that electroweak radiative corrections can take them outside the direct exclusion limit for technimesonic pseudoscalars \cite{Dietrich:2009ix}, which is slightly above the mass of the weak gauge bosons \cite{Amsler:2008zzb}. On top of that, these extra Nambu--Goldstone-modes carry technibaryon number and hence, as opposed to technimesons, they can only be produced in pairs. The positive electroweak contributions to the squared masses of the Nambu--Goldstone-modes also stabilise the embedding of the electroweak gauge group, which leads to its correct breaking \cite{Peskin:1980gc}. It could have been embedded in such a way that it remains unbroken, but the latter embedding would be destabilised by electroweak radiative corrections.

\begin{figure}[t]
\resizebox{8.5cm}{!}{\includegraphics{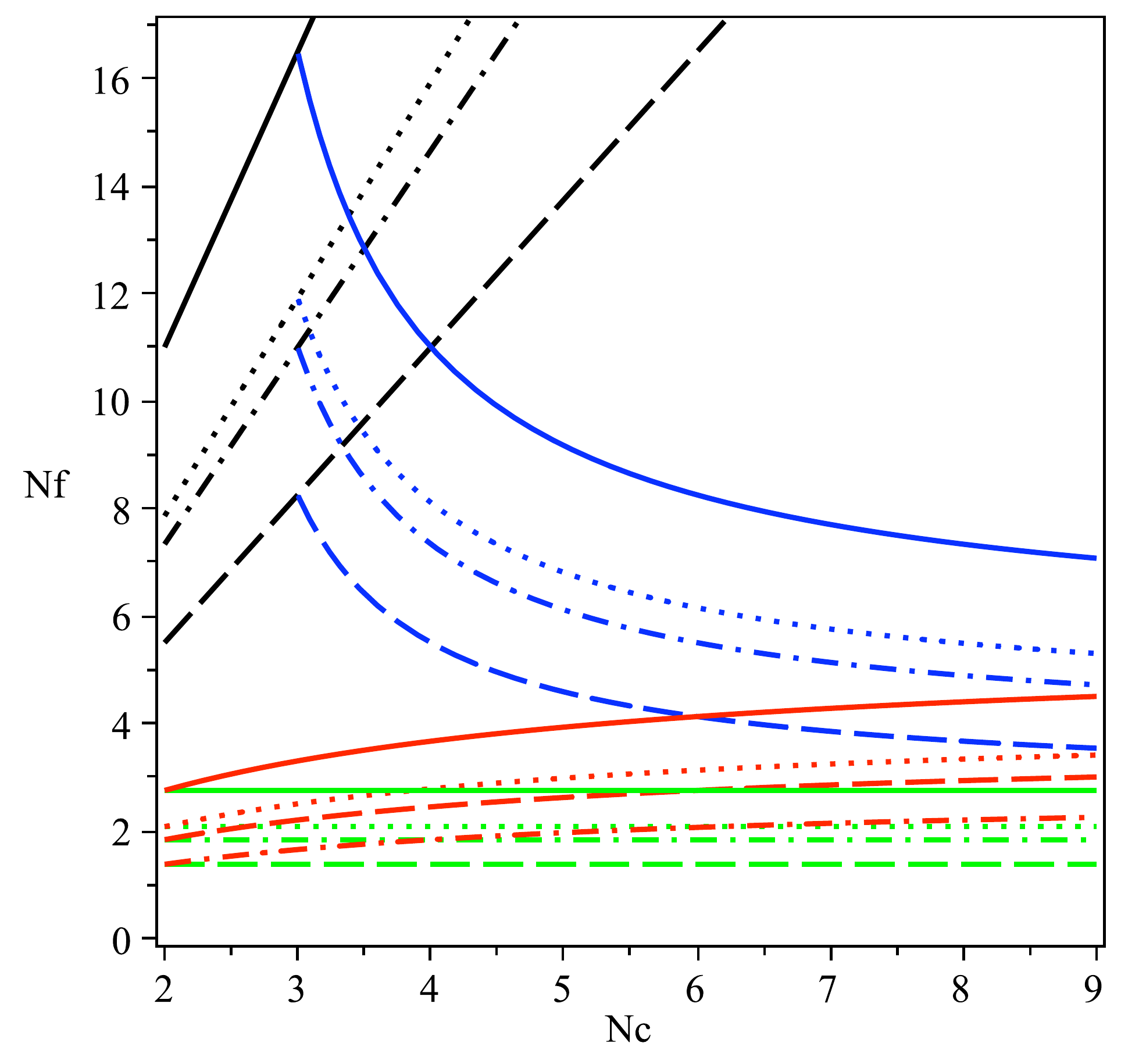}}
\caption{Massless fermions:
Conformal window for $SU(N)$ gauge theories with fermions in the (from top to bottom) fundamental (black; straight, rising), two-index antisymmetric (blue; curved, falling), two-index symmetric (red; curved, rising) or adjoint (green; straight, horizontal) representation of the gauge group. Above the solid curves asymptotic freedom is lost. The dotted lines show the lower bound for the conformal window according to the rainbow-ladder-approximation to the Dyson--Schwinger-equations. The lower bounds for the conformal window are depicted by the dash-dotted ($\gamma=1$) and the dashed ($\gamma=2$; minimal lower bound) curves.}
\label{pd}
\end{figure}

\begin{figure}[t]
\resizebox{8.5cm}{!}{\includegraphics{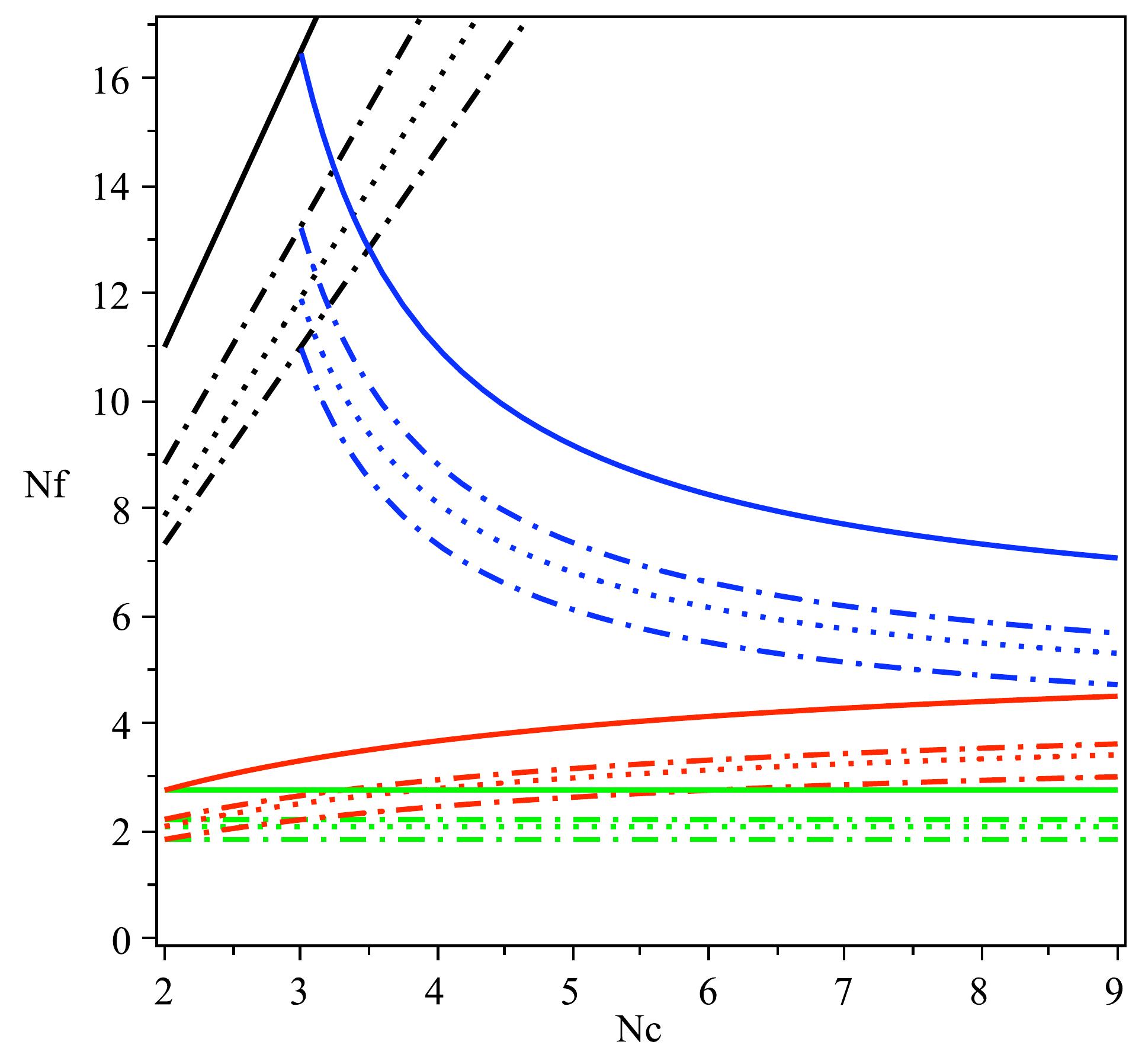}}
\caption{Massive fermions, $\gamma=1$:
Quasiconformal window for $SU(N)$ gauge theories with fermions in the (from top to bottom) fundamental (black; straight, rising), two-index antisymmetric (blue; curved, falling), two-index symmetric (red; curved, rising) or adjoint (green; straight, horizontal) representation of the gauge group. Above the solid curves asymptotic freedom is lost. The dotted lines show the lower bound for the conformal window according to the rainbow-ladder-approximation to the Dyson--Schwinger-equations. For massless fermions the lower bound is depicted by the lower dash-dotted curve; for fermions with $x=-4$ ($\mu=4m$), by the upper dash-dotted curve.}
\label{pd1}
\end{figure}

\begin{figure}[t]
\resizebox{8.5cm}{!}{\includegraphics{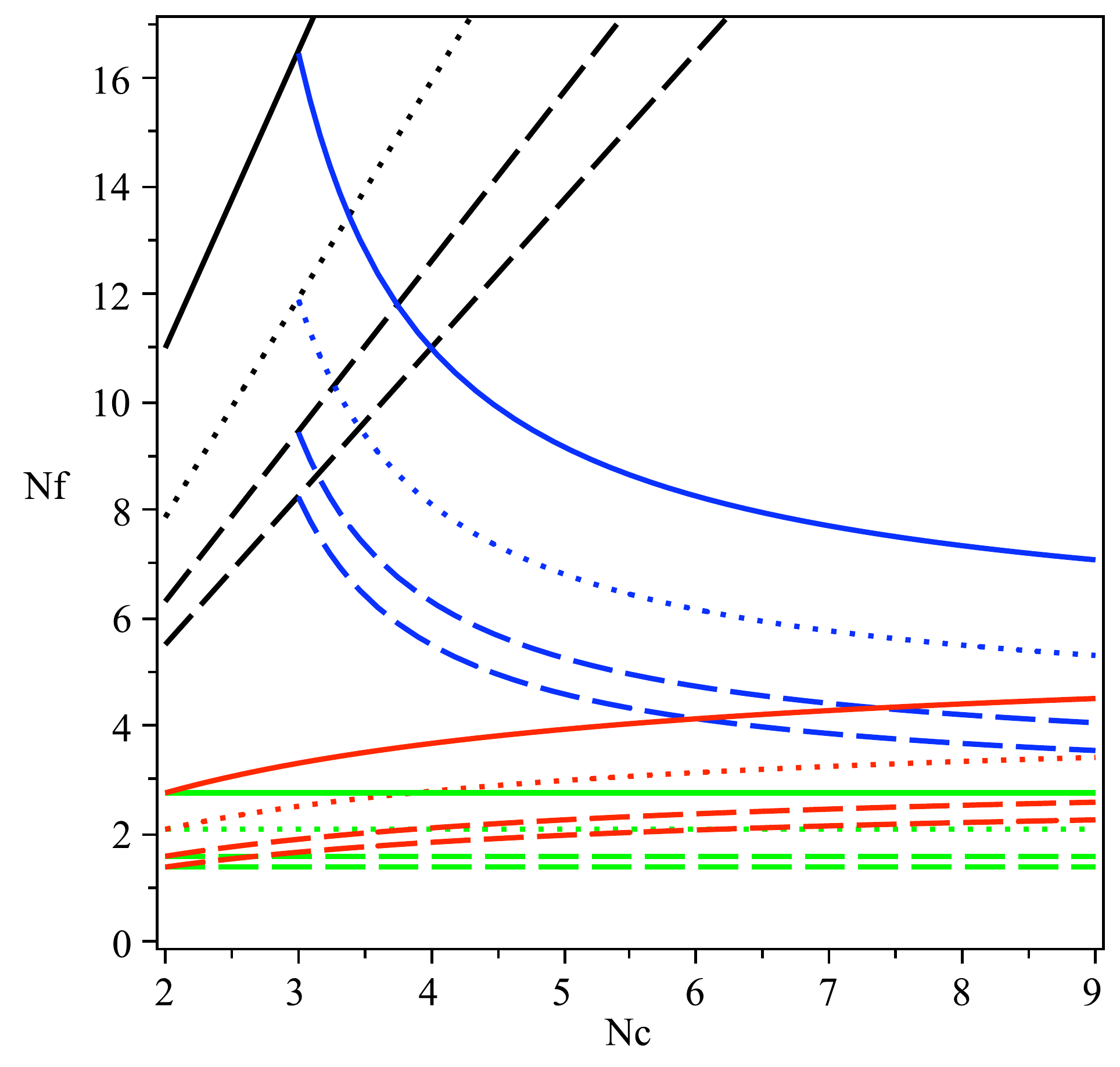}}
\caption{Massive fermions, $\gamma=2$:
Quasiconformal window for $SU(N)$ gauge theories with fermions in the (from top to bottom) fundamental (black; straight, rising), two-index antisymmetric (blue; curved, falling), two-index symmetric (red) or adjoint (green; straight, horizontal) representation of the gauge group. Above the solid curves asymptotic freedom is lost. The dotted lines show the lower bound for the conformal window according to the rainbow-ladder-approximation to the Dyson--Schwinger-equations. For massless fermions the lower bound is depicted by the lower dashed curve; for fermions with $x=-4$ ($\mu=4m$), by the upper dashed curve.}
\label{pd2}
\end{figure}

\begin{figure}[t]
\resizebox{8.5cm}{!}{\includegraphics{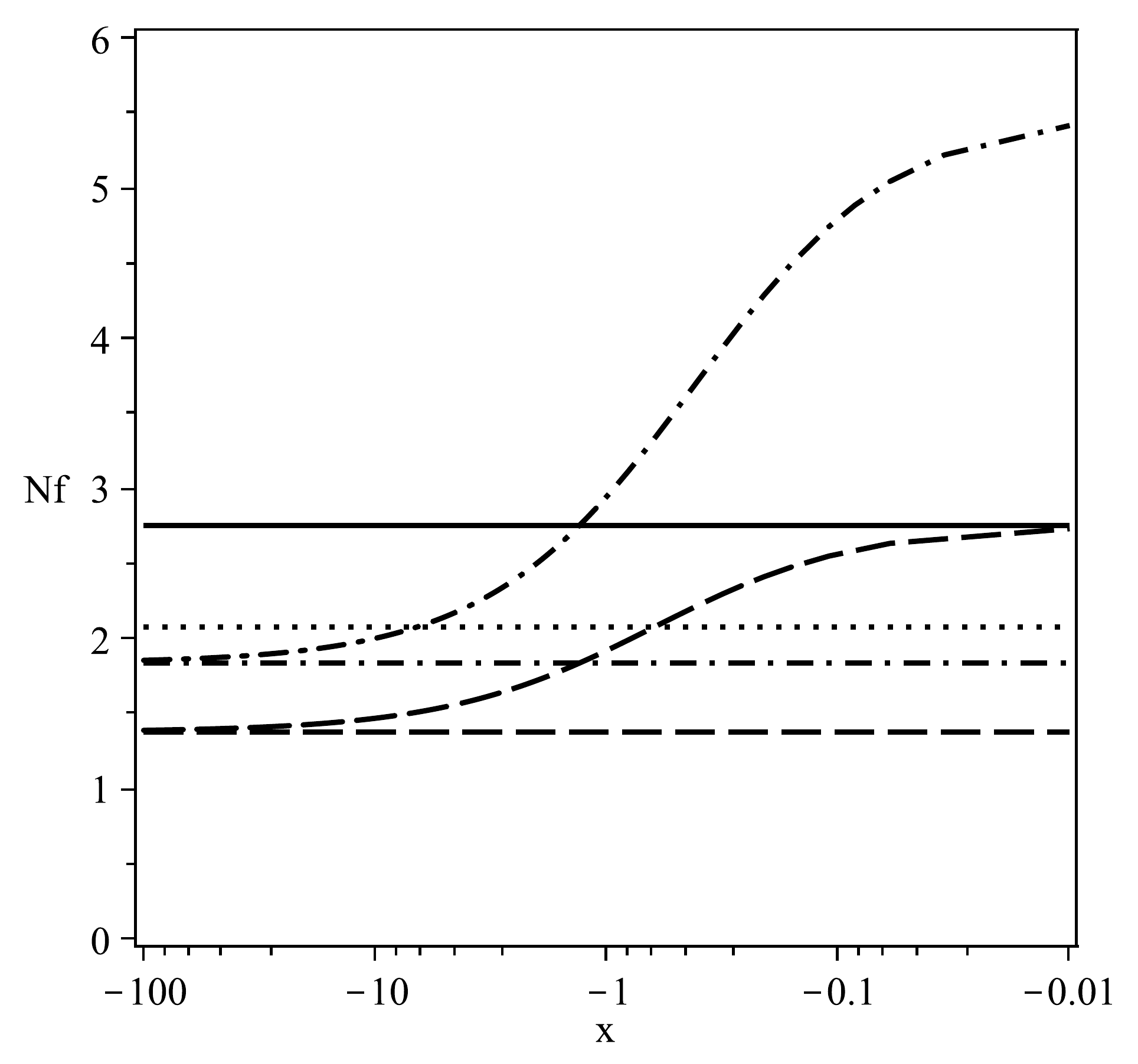}}
\caption{Lower bounds for the (quasi)conformal window, two flavours, adjoint representation, $SU(2)$:
ladder-rainbow-approximation (dotted);
$\gamma=1$ massless (horizontal straight line, dash-dotted), massive (curve, dash-dotted);
$\gamma=2$ massless (horizontal straight line, dashed), massive (curve, dashed).
Above the solid line asymptotic freedom is lost.
The zone in which the flavours are gradually switched off, spans four orders of magnitude.
}
\label{mwt}
\end{figure}

\begin{figure*}[t]
\resizebox{16.4cm}{!}{\includegraphics{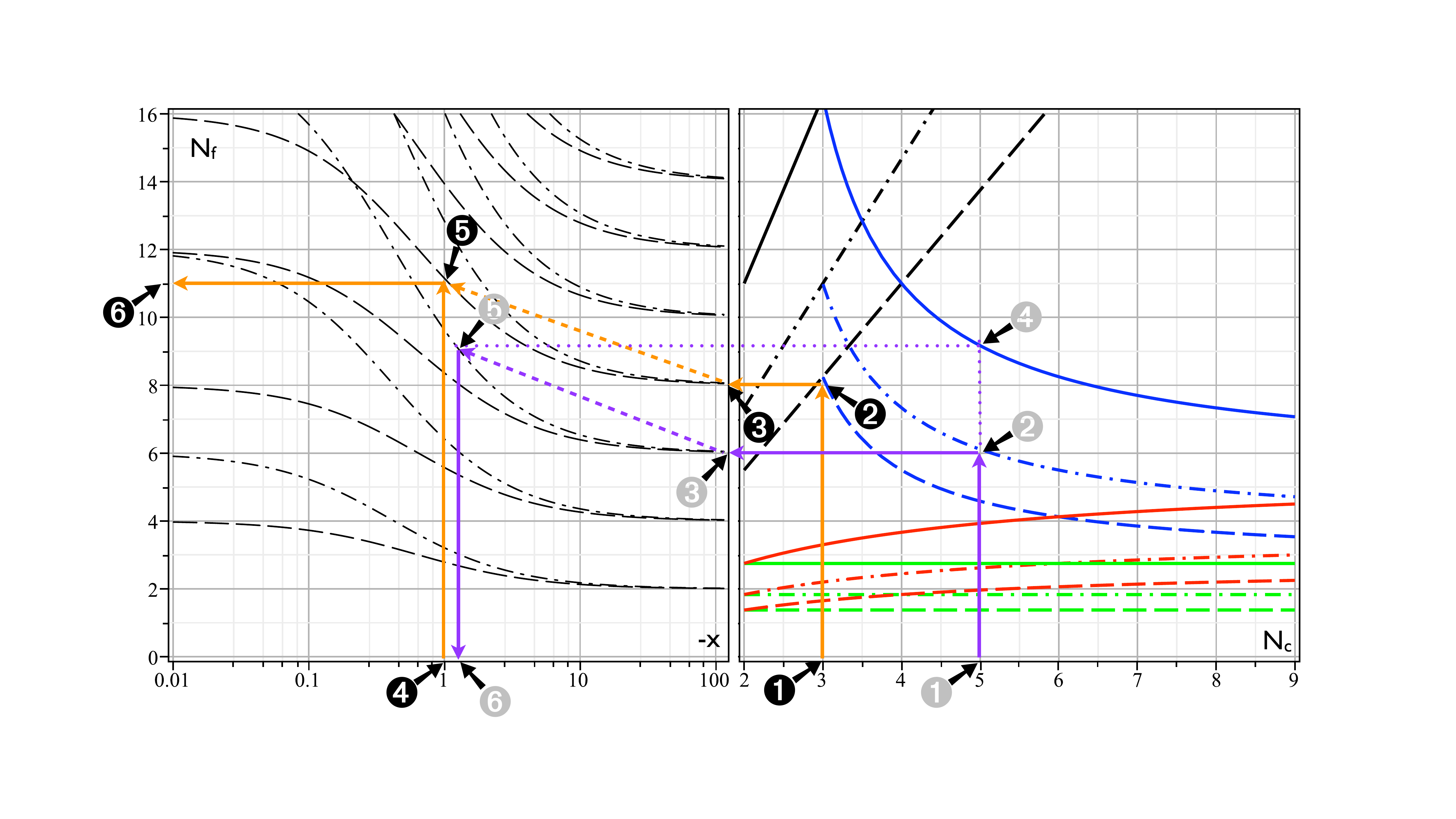}}
\caption{
Examples for how to use Figs.~\ref{su}, \ref{sp} and \ref{so} on the example of the (quasi)conformal window for $SU(N_c)$ theories. More explanations are given at the end of the body of the text of the Appendix.
}
\label{howto}
\end{figure*}

This is different in the pseudoreal case. There, the breaking goes to $Sp(4)$, which yields two uneaten Nambu-Goldstone-modes. They receive negative contributions to their squared masses from electroweak radiative corrections. Thus, the embedding that leads to the correct breaking of the electroweak gauge group is destabilised. It must be stabilised by an additional mechanism, which is usually incorporated with the extended technicolour sector. In fact, looking at the non-(pseudo)real case, all it takes to provide suitable masses to the weak gauge bosons is an $SU(2)_L\times SU(2)_R\rightarrow SU(2)_V$ breaking pattern. Hence, above the technicolour scale not the full $SU(4)$ symmetry has to be preserved, but only the $SU(2)_L\times SU(2)_R$.

Looking, for the moment, exclusively at the two-flavour technicolour sector neither with electroweak interactions nor extended technicolour effects, it consists of four Weyl fermions, $U_L$, $D_L$, $U_R$, and $D_R$. Representing the right-handed fields such that they transform as left-handed fields, $-i\sigma^2U_R^*$ and $-i\sigma^2D_R^*$, the $SU(4)$ flavour symmetry becomes apparent. The assignment of the left and right fields is dictated by the coupling to the electroweak interactions. From the viewpoint of the flavour symmetries, the mass terms 
\begin{equation}
\mathcal{L}_m = m~(\bar{U}_LU_R + \bar{U}_RU_L + \bar{D}_LD_R + \bar{D}_RD_L)
\end{equation}
and
\begin{equation}
\mathcal{L}_\lambda= \lambda~(\bar{U}_LD_L + \bar{D}_LU_L + \bar{U}_RD_R + \bar{D}_RU_R)
\end{equation}
are equivalent: Imagine, for example, that the four Weyl flavours are arranged in a column vector. Then the respective mass terms can be expressed by contraction with the matrices
\begin{equation}
\left(\begin{array}{cc}\mathbbm{O}&\mathbbm{1}\\\mathbbm{1}&\mathbbm{O}\end{array}\right)
\mathrm{~~~or~~~}
\left(\begin{array}{cc}\mathbbm{1}&\mathbbm{O}\\\mathbbm{O}&-\mathbbm{1}\end{array}\right),
\end{equation}
which are linked by an $SU(4)$ transformation. $\mathbbm{O}$ and $\mathbbm{1}$ are $2\times2$ zero and unit matrices, respectively.

The first mass term above, however, breaks the electroweak symmetry, while the second does not; it breaks the $SU(4)$ to $SO(4)\simeq SU(2)_L\times SU(2)_R$. Consequently, it contributes to the masses of those Nambu--Goldstone-modes, which link left- with right-fields, that is, the modes with non-zero technibaryon number. Apart from a direct application to dynamical electroweak symmetry breaking, for which the electroweak symmetry must be unbroken before the chiral condensate is formed, the investigation of the impact of an explicit mass term of $\mathcal{L}_m$-type is interesting per se as well as for quantum chromodynamics, and natural for a study in the framework of lattice field theory.

For techniquarks in a pseudoreal representation of the technicolour gauge group, terms which break the $SU(4)$ flavour symmetry to $SU(2)_L\times SU(2)_R$ are needed to stabilise the vacuum. The motivation for studying additional mass terms of $\mathcal{L}_\lambda$-type for real representations (and yet another motivation for studying it for pseudoreal representations) is to control the amount of walking or even avoid conformality of an otherwise promising candidate \cite{Hietanen:2008mr}. (In fact, it appears to be interesting to compare the implications from our mass-dependent all-order $\beta$-function with results from lattice studies, which is an active field in the context of walking technicolour theories \cite{Hietanen:2008mr,Catterall:2007yx}.) For this purpose, we do not even have to increase the value of the mass parameter $\lambda$ beyond its value when used to make the extra Nambu--Goldstone-modes sufficiently massive: The extended-technicolour-induced term is of dynamical origin. Thus, arguing based on a Gell-Mann--Oakes--Renner relation,
\begin{equation}
m_\pi^2f_\pi^2=2\lambda\langle Q\bar{Q}\rangle ,
\end{equation}
where $m_\pi=O(\mathrm{few}~0.1 \mathrm{TeV})$ is the extended technicolour contribution to the mass of the corresponding pions, $f_\pi=O(\mathrm{few~TeV})$ their decay constant and $\langle Q\bar{Q}\rangle=O(\mathrm{TeV}^3)$ the related techniquark condensate, results in $\lambda$ being of $O(\mathrm{TeV})$ as well. This is where also the technicolour scale is situated. Hence, an interference of related threshold effects with the technicolour phase transition appears natural. Thus, the extended technicolour might
affect the technicolour phase transition.

On the other hand, the dynamically generated techniquark mass in the $\mathcal{L}_m$ channel can be estimated to be \cite{Appelquist:1990wr}
\begin{equation}
\Sigma(0)\approx 2\pi F_\pi/\sqrt{d_\mathrm{R}}, 
\end{equation}
where for technicolour models with two techniflavours $F_\pi=\Lambda_\mathrm{ew}=246\mathrm{GeV}$. 
$d_\mathrm{R}$ is the dimension of the representation of the technicolour gauge group with respect to which the techniquarks transform. Hence, $\Sigma(0)$ is also of $O(\mathrm{TeV})$ and it seems likely that also the critical value of the coupling $\alpha^*$ is influenced by the presence of $\mathcal{L}_\lambda$. 

As another example consider partially (electroweakly) gauged technicolour \cite{Dietrich:2005jn}. In partially gauged technicolour, be it with matter in a single or different simultaneously present representations of the technicolour gauge group, only some of the techniquarks are gauged under the electroweak gauge group. This is usually done to alleviate constraints from electroweak precision data, in particular, from the oblique parameters. At the same time it allows to bring the theory close to the (quasi)conformal window. The part of the techniquarks that is not gauged under the electroweak can be given masses without breaking the electroweak symmetry. Depending on the flavour symmetry among the electroweakly gauged techniquarks of a partially gauged technicolour model, specific mass terms that leave the electroweak symmetry intact might be possible in this sector as well, as was already explained above.

\begin{figure*}[t!]
\resizebox{16cm}{!}{\includegraphics{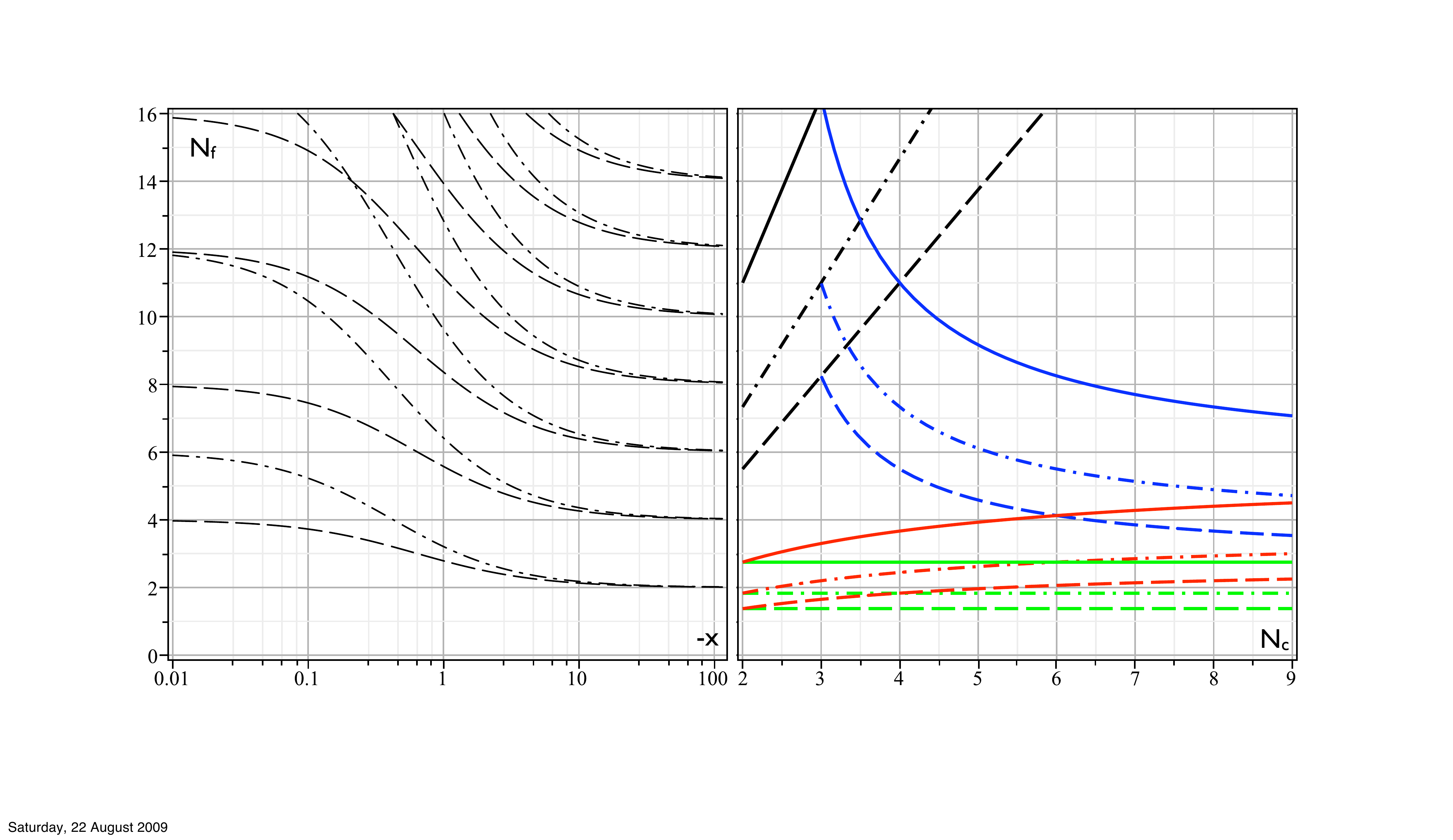}}
\caption{
(Quasi)conformal window for $SU(N_c)$ theories. 
\underline{Right-hand side}:
Conformal window for $SU(N_c)$ gauge theories with fermions in the (from top to bottom) fundamental (black; straight, rising), two-index antisymmetric (blue; curved, falling), two-index symmetric (red; curved, rising) or adjoint (green; straight, horizontal) representation of the gauge group. Above the solid curves asymptotic freedom is lost. The lower bounds for the conformal window are depicted by the dash-dotted ($\gamma=1$) and the dashed ($\gamma=2$; minimal lower bound) curves.
\underline{Left-hand side}:
Effective rescaling of the $N_f$ axis due to massive fermions for $\gamma=1$ (dash-dotted) and $\gamma=2$ (dashed).
}
\label{su}
\end{figure*}

\begin{figure*}[t]
\resizebox{16cm}{!}{\includegraphics{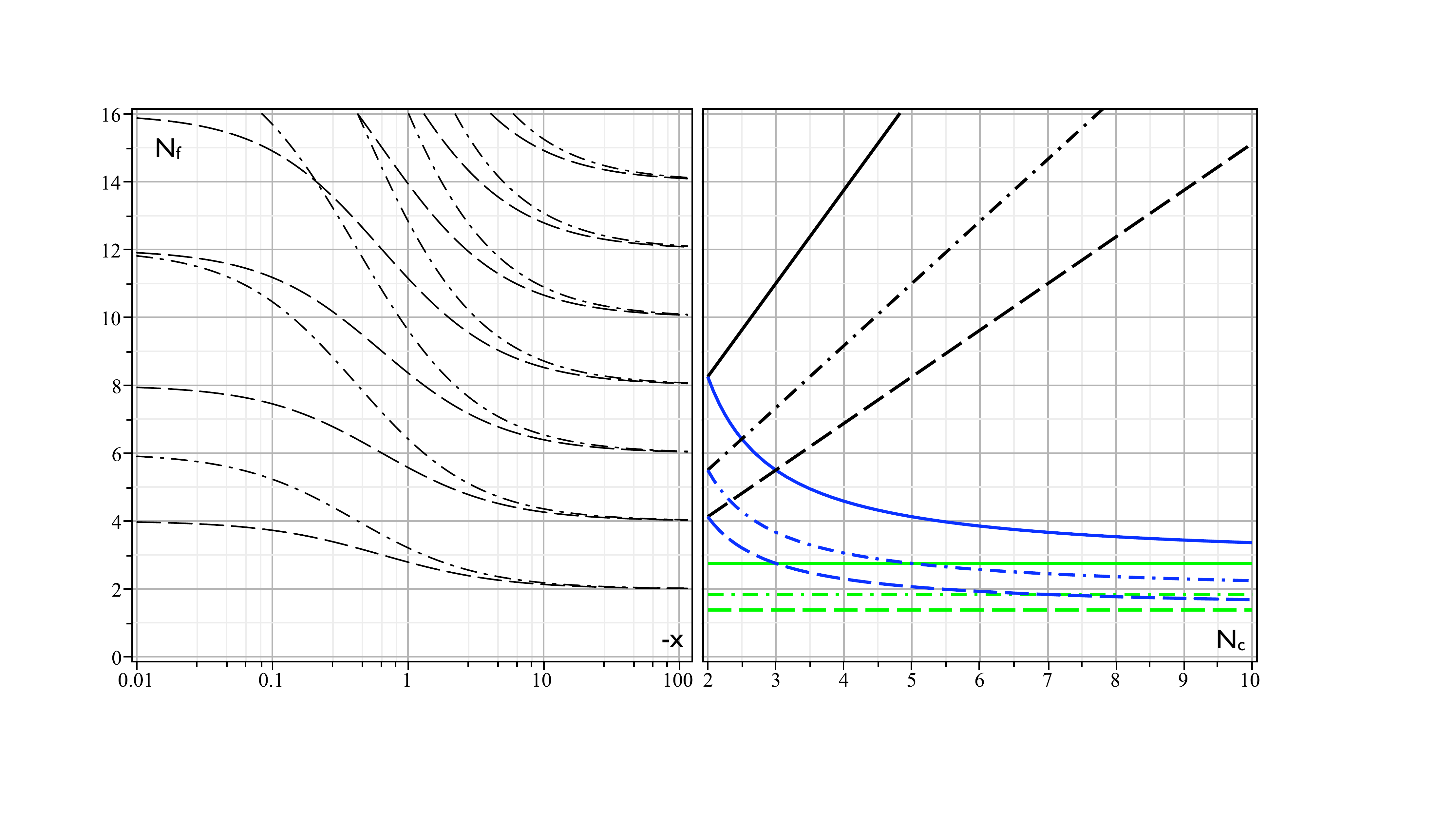}}
\caption{
(Quasi)conformal window for $Sp(2N_c)$ theories.
\underline{Right-hand side}:
Conformal window for $Sp(2N_c)$ gauge theories with fermions in the (from top to bottom) fundamental (black; straight, rising), two-index antisymmetric (blue; curved, falling) or adjoint (green; straight, horizontal) representation of the gauge group. Above the solid curves asymptotic freedom is lost. The lower bounds for the conformal window are depicted by the dash-dotted ($\gamma=1$) and the dashed ($\gamma=2$; minimal lower bound) curves.
\underline{Left-hand side}:
Effective rescaling of the $N_f$ axis due to massive fermions for $\gamma=1$ (dash-dotted) and $\gamma=2$ (dashed).
}
\label{sp}
\end{figure*}

\begin{figure*}[t]
\resizebox{16cm}{!}{\includegraphics{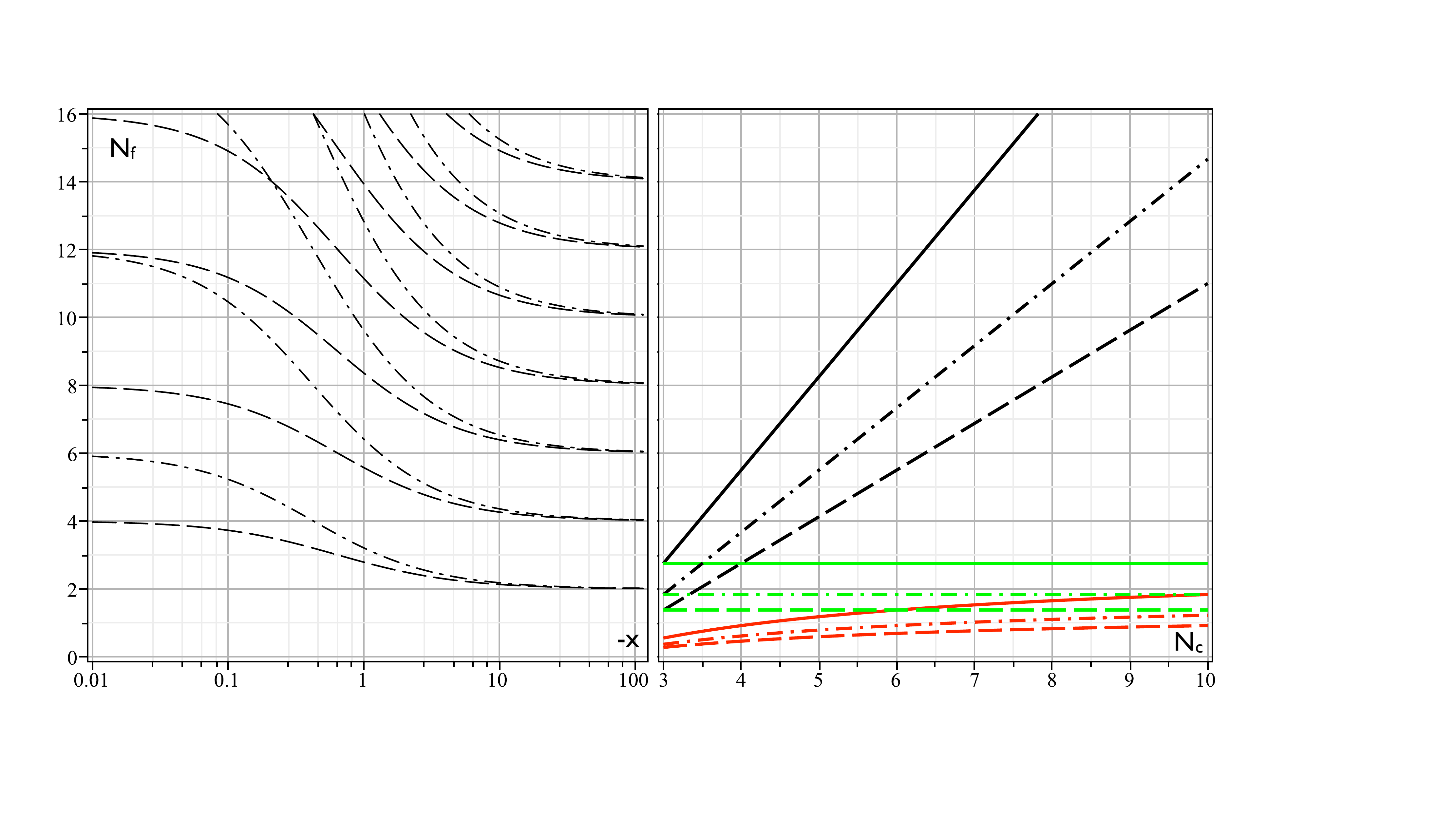}}
\caption{(Quasi)conformal window for $SO(N_c)$ theories.
\underline{Right-hand side}:
Conformal window for $SO(N_c)$ gauge theories with fermions in the (from top to bottom) fundamental (black; straight, rising), two-index symmetric (red; curved, rising) or adjoint (green; straight, horizontal) representation of the gauge group. Above the solid curves asymptotic freedom is lost. The lower bounds for the conformal window are depicted by the dash-dotted ($\gamma=1$) and the dashed ($\gamma=2$; minimal lower bound) curves.
\underline{Left-hand side}:
Effective rescaling of the $N_f$ axis due to massive fermions for $\gamma=1$ (dash-dotted) and $\gamma=2$ (dashed).
}
\label{so}
\end{figure*}


\subsection{The conformal window\label{CONF2}}

In \cite{Dietrich:2006cm}, the lower bound of the conformal window in the plane spanned by the number of colours of an $SU(N)$ theory and the number of flavours was determined by equating the coupling at the Caswell--Banks--Zaks fixed point with the critical coupling \cite{Appelquist:1988yc} for the formation of a chiral condensate as obtained within the ladder-rainbow-approximation to the Dyson--Schwinger--equations. Solving for the number of flavours as a function of the number of colours yields the dotted curves in Fig.~\ref{pd}, for four different representations; from top to bottom, the fundamental, the two-index antisymmetric, the two-index symmetric and the adjoint. Asymptotic freedom is lost above the solid curves.

In \cite{Ryttov:2007cx}, the conjectured $\beta$-function (\ref{fsaob}) was used to determine the lower bound of the conformal window. To this end, it was equated to zero and the value of the anomalous dimension $\gamma$ was kept fixed. While the ladder-rainbow-approximation yields a critical value of one, the only theoretically hard upper bound on $\gamma$ arises from the requirement of unitarity of the gauge field theory and is two. (This follows from the fact that in a conformal field theory the dimension $3-\gamma$ of all non-trivial spinless operators and thus, that of the chiral condensate must be larger or equal to unity \cite{Mack:1975je}.) The lower bound for the conformal window must, hence, not lie below the thus obtained curve. (There are also indications from duality arguments \cite{Sannino:2009qc} for choosing the maximum value for $\gamma$.) Again, after solving for the number of flavours as a function of the number of colours, this approach gives rise to the dash-dotted curves ($\gamma=1$) and dashed curves ($\gamma=2$), respectively, in Fig.~\ref{pd}. This approach has also been applied to gauge groups other than $SU(N)$ and to multiple representations \cite{Sannino:2009aw,Ryttov:2009yw}.
Interestingly, this approach leads to a universal formula
$
1=\kappa~2N_f~T(R)/C_2(G)
$
for the lower bound of the conformal window \cite{Armoni:2009jn}. In \cite{Armoni:2009jn}, this relation was found in the framework of the worldline formalism, and $\kappa\approx 1/4$ was determined from matching to SQCD. From Eq.~(\ref{fsaob}), one finds $\kappa=(2+\gamma)/11$. (A combination of the two results would yield $\gamma\approx 3/4$.)
{
Such a relation was also found in \cite{Poppitz:2009uq}.
}

Here, we study the influence of threshold effects due to the fermion masses based on the $\beta$-function (\ref{maob}) by fixing $\bar{\gamma}$ to the two benchmark values from the massless study. For vanishing mass, we find the same results as before. For non-zero values of the mass,
the ``lower bound of the conformal window" is moved towards a larger number of flavours. 
In the presence of massive fermions this term (``\dots") is somewhat abused, here. Clearly, in a theory, where all fermions have a nonzero mass, however tiny it may be, they freeze out for scales far enough below this mass, and we are effectively left with a pure Yang--Mills theory, where the antiscreening from the gluons is unchallenged.
Hence, in a theory in which all flavours are massive, there is strictly speaking no conformal window. Therefore, what above was called ``lower bound of the conformal window" in the massive case is the phenomenologically decisive minimal number of flavours above which the coupling develops a plateau (walks). Consequently, we should talk of a ``quasiconformal window", which in the limit $m\rightarrow 0$ coincides with the conformal window. In the massless case, the walking setups were to be found slightly below the lower edge of the conformal window. (There, above the edge, the theory does evolve into the fixed point.) In the massive case, in a setup slightly below the modified bound, there will also be at least some walking. In that case, the amount of walking, that is, the range of scales of quasiconformal behaviour, is determined by a direct interplay of the freezing out of the flavours due to the explicit mass and the onset of chiral symmetry breaking. (Independent of walking, a similar interplay between quark mass effects and chiral symmetry breaking exists in quantum chromodynamics for the strange quark.) If the number of flavours is really below this bound, it never comes near the fixed point and does not show any walking. If the number of flavours is above this bound (and the theory is still asymptotically free) the theory approaches the fixed point very closely and stays in its vicinity until the flavours start decoupling gradually. Once the flavours are decoupled sufficiently, the coupling starts running again. This means also that in such a setting, the position of the low-energy end of the plateau is not determined by the initial conditions for the renormalisation group evolution, but by the value of the fermion masses. 
As examples we present the quasiconformal windows for $\gamma=1$ and $x=-4$ ($\mu=4m$) in Fig.~\ref{pd1} and for $\gamma=2$ and $x=-4$ in Fig.~\ref{pd2}. For comparison, the massless case is also always shown. (We concentrate here on equal masses for all fermions. The phenomenology becomes richer, once we allow for different masses, which, on the other hand, would also break more symmetries.)

At this point, let us revisit briefly the notion of ``number of active flavours". The determination of the conformal window relies on setting the $\beta$-function (\ref{maob}) equal to zero. This amounts to finding the zero of its numerator. For fixed $\bar{\gamma}$, the latter depends only on $\bar{\beta}_0$ and not $\bar{\beta}_1$. Hence, 
the number of active flavours that counts, at least for this way of determining the quasiconformal window, is $N_{f,0}$. This aspect together with the fact that $N_{f,1}$ is not monotonously interpolating between the limiting cases indicates that $N_{f,1}$ or $\bar{\beta}_1$, and actually probably all the $N_{f,j}$ or $\bar{\beta}_j$ with $j\ge1$, encode something else than a mere modification of the number of flavours.

In Fig.~\ref{mwt}, in order to show the influence of the mass of the fermions, we plot the critical number of flavours as a function of $x=-\mu^2/(4m^2)$ for an $SU(2)$ field theory with two adjoint Dirac flavours, that is, the core of minimal walking technicolour. First of all, one can see that the switching zone spans four orders of magnitude. In other words, threshold effects are felt for energy scales, which are a hundred times bigger than the mass of the fermions. Further, taking the mass to infinity ($x$ to zero) does not lead to the zero flavour result because $\gamma$ is kept fixed. As for the range of the mass effect, for $\gamma=1$, an $x$ of $O(-10^1)$ lifts the lower bound to the value obtained from the ladder-rainbow-approximation. A value for $x$ slightly below $-1$, that is, $2m\lesssim\mu$, closes the quasiconformal window altogether, as asymptotic freedom is lost, before quasiconformality is reached. For $\gamma=2$, the ladder-rainbow-value is only reached for $x$ slightly above $-1$, that is, $2m\gtrsim\mu$; the quasiconformal window closes for infinite fermion masses, $x\rightarrow 0$.

In principle, the quasiconformal windows, shown in Figs. \ref{pd1} and \ref{pd2} for $SU(N_c)$ gauge theories, have to be replotted for every value of $x$. Therefore, in the Appendix A, we have attempted to give a universal representation that permits to read off $N_f$ for the lower bound of the quasiconformal window for $SU(N_c)$ (Fig.~\ref{su}), $Sp(2N_c)$ (Fig.~\ref{sp}) and $SO(N_c)$ (Fig.~\ref{so}) gauge theories, from single plots. Fig.~\ref{howto} explains the procedure.


\section{Summary\label{SUMM}}

We propose a mass-dependent all-order $\beta$-function (\ref{maob}), which combines the ideas that, in \cite{Ryttov:2007cx}, led to the supersymmetry inspired mass-independent all-order $\beta$-function (\ref{fsaob}) 
with results for the $\beta$-function coefficients in the mass-dependent background field momentum subtraction scheme. 
(Formally, it is legal to use mass-independent subtraction schemes. They, however, neglect completely the influence of the constituent masses on the evolution of the gauge 
coupling constant.)
For vanishing fermion masses, the mass-independent all-order $\beta$-function (\ref{fsaob}) is recovered from the mass-dependent all-order $\beta$-function (\ref{maob}). We use the $\beta$-function (\ref{maob}) to study the impact of explicit fermion masses on the lower bound of the (quasi)conformal window of the corresponding field theory: Effects of the fermion mass can be sensed at scales more than a hundred times bigger than the fermion mass; (See Fig.~\ref{mwt}.) the lower bound of the (quasi)conformal window can be moved to a considerably higher number of flavours; in fact, it can be closed up altogether. (See Figs.~\ref{pd1}, \ref{pd2} and \ref{mwt}.) 
This means that the corresponding theory does not show quasiconformal behaviour for any number of flavours below the value for which asymptotic freedom is lost.
In the investigation of dynamical electroweak symmetry breaking by walking technicolour theories, our approach may be useful for studying effects of extended technicolour interactions, which may be primarily required to stabilise the vacuum and/or render additional Nambu--Goldstone-modes sufficiently massive. (These effects interfere with the chiral dynamics of technicolour in a similar way as electroweakly induced masses interfere with the chiral dynamics of quantum chromodynamics.) Additionally, it might be extended technicolour effects that make a theory quasiconformal, which from its bare technicolour structure would be completely conformal and hence, not suited for breaking the electroweak symmetry dynamically.
It would be interesting to address the issue of mass dependence, in particular, and of other extended technicolour effects, in general, in lattice simulations and to contrast them to the present investigation.


\section*{Acknowledgments}

The author would like to thank
A.~Armoni,
R.~Barbieri,
S.~J.~Brodsky,
M.~T.~Frandsen,
M.~J\"arvinen,
and
F.~Sannino
for discussions.
The work of DDD was supported by the Danish Natural Science Research Council.



\appendix

\section{Quasiconformal windows}

The critical number of flavours obtained by setting the mass-dependent $\beta$-function equal to zero at a fixed value for $\bar{\gamma}$ is given by,
\begin{equation}
N_f=\frac{11}{2}\frac{C_2(G)}{T(R)}[\bar{\gamma}+2b_0(x)]^{-1} .
\end{equation}
The modification due to the mass of the fermion is, hence, universal in the sense that it neither depends on the gauge group nor the representation. The latter is encoded in the fraction 
$C_2(G)/T(R)$, while the mass effect are contained in $b_0(x)$. 

We are here making use of this universality by giving an illustration for the quasiconformal windows for theories constructed from the fundamental, 2-index antisymmetric, 2-index symmetric or adjoint representation of an $SU(N_c)$ (Fig.~\ref{su}), $Sp(2N_c)$ (Fig.~\ref{sp}) or $SO(N_c)$ (Fig.~\ref{so}) gauge group. \{The $m=0$ case was first treated in \cite{Dietrich:2006cm} ($SU$), \cite{Sannino:2009aw} ($Sp$, $SO$)\}. Said representation consists of split figures; their left half depicts the quasiconformal windows for vanishing mass and their right half the modification due to a nonzero mass, for both benchmark values, $\bar{\gamma}=1$ and $\bar{\gamma}=2$, respectively. 

Fig.~\ref{howto} explains how to use the depiction:

On the right-hand side, we have the
conformal window for $SU(N_c)$ gauge theories with fermions in the (from top to bottom) fundamental (black; straight, rising), two-index antisymmetric (blue; curved, falling), two-index symmetric (red; curved, rising) or adjoint (green; straight, horizontal) representation of the gauge group [here $SU(N_c)$]. Above the solid curves, asymptotic freedom is lost. The lower bounds for the conformal window are depicted by the dash-dotted ($\gamma=1$) and the dashed ($\gamma=2$; minimal lower bound) curves.

The left-hand side shows the 
effective rescaling of the $N_f$ axis due to massive fermions for $\gamma=1$ (dash-dotted) and $\gamma=2$ (dashed).
Note that the curves signalling the loss of asymptotic freedom are not rescaled.

The black numbers mark the first example, in which we read off the lower bound of the quasiconformal window for $x=-1$ for fermions in the fundamental representation of $SU(3)$ with the criterion $\gamma=2$: 
1) In the plot on the right-hand side, pick $N_c=3$;
2) go up to the $\gamma=2$ curve for fermions in the fundamental representation, which yields $N_f\approx 8$ for the lower bound of the massless conformal window;
3) find the $N_f=8$ curve in the plot on the left-hand side;
4) pick the $-x=1$ line;
5) find the crossing between the curve and the line;
6) read off the result on the axis; here, $N_f\approx 11$.

The grey numbers indicate the second example, in which
the value of $-x$ is to be determined for which the quasiconformal window closes up completely for fermions in the 2-index antisymmetric representation of $SU(5)$ and using the criterion $\gamma=1$:
1) In the plot on the right-hand side, pick $N_c=5$;
2) go to the $\gamma=1$ curve for fermions in the 2-index antisymmetric representation, which yields $N_f\approx 6$ for the lower bound of the massless conformal window;
3) find the $N_f=6$ curve in the plot on the left-hand side;
4) determine the number of flavours for which asymptotic freedom is lost from the plot on the right-hand side, which is $N_f\gtrsim 9$;
5) find where the curve in the plot on the left-hand side reaches this value;
6) go down to the $-x$ axis, which yields $-x\gtrsim 1$.


\end{document}